\documentclass[preprint2]{aastex}
    
\usepackage{epsfig}
\usepackage{rotating}

\shorttitle{Detecting planets in protoplanetary disks}
\shortauthors{Wolf S.\ et al.}

\begin{document}


\title{Detecting planets in protoplanetary disks: A prospective study}


\author{S. Wolf}
\affil{Th\"uringer Landessternwarte Tautenburg, 
  Sternwarte 5, D--07778 Tautenburg, Germany}
\email{wolf@tls-tautenburg.de}

\author{F. Gueth}
\affil{Institut de Radio Astronomie Millim\'etrique, 300 
  rue de la Piscine, 38406 Saint Martin d'H\`eres, France}
\email{gueth@iram.fr}

\author{Th. Henning}
\affil{Astrophysikalisches Institut und Universit\"ats-Sternwarte, 
  Schillerg\"asschen 2-3, D--07745 Jena, Germany}
\email{henning@astro.uni-jena.de}

\and

\author{W. Kley}
\affil{Universit\"at T\"ubingen, Inst.\ f\"ur Astronomie und Astrophysik,
Abt. Computational Physics, Auf der Morgenstelle 10,
D-72076 T\"ubingen, Germany}
\email{kley@tat.physik.uni-tuebingen.de}

\begin{abstract}
We investigate the possibility to find evidence for planets in circumstellar disks
by infrared and submillimeter interferometry.
We present simulations of a circumstellar
disk around a solar-type star with an embedded planet of 1 Jupiter mass.
The three-dimensional (3D) density structure of the disk results from hydrodynamical
simulations. On the basis of 3D radiative transfer
simulations, images of this system were calculated. The intensity maps provide
the basis for the simulation of the interferometers VLTI (equipped with 
the mid-infrared instrument MIDI) and ALMA.
While MIDI/VLTI will not provide the possibility to distinguish between
disks with or without a gap on the basis of visibility measurements, 
ALMA will provide the necessary basis for a direct gap detection. 
\end{abstract}



\section{Introduction}\label{intro}

Based on studies of the evolution of protoplanets in protoplanetary disks,
it has been established that - depending on the hydrodynamic properties
of the planet and the disk - giant protoplanets may open a gap and cause spiral
density waves in the disk (see, e.g., Kley 1999, Kley et al.\ 2001). 
The gap may extend up to several AU in width.
Thus, the question arises if one can find such a gap as an indicator 
for the presence of a protoplanet with present-day or near-future techniques.
In order to study this possibility, we use hydrodynamical simulations of
a protoplanetary disk with an embedded planet (\S~\ref{model}) and compute
the expected brightness distributions, with a 3D radiative transfer code
(\S~\ref{rttech}). The resulting images are presented in \S~\ref{images}.
Finally, we show that ALMA will provide the necessary basis to detect gaps 
in circumstellar disks in the mm/submm wavelength range (\S~\ref{simmial}).

\section{The disk model}\label{model}

\begin{figure}[h]
  \epsscale{1.0}
  \plotone{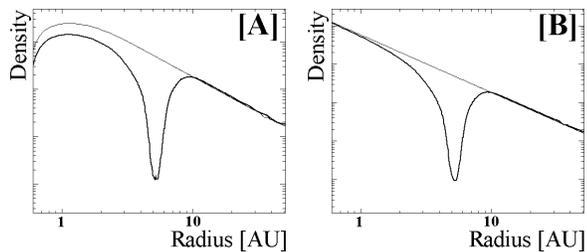}
  \caption{Radial density profile in the midplane of the disk
    (averaged in $\phi$ direction, arbitrary scale). The thin lines
    represent the corresponding density distributions without an embedded planet.
    [A] With accretion onto the central object (open inner disk boundary).
    [B] No accretion onto the central object (closed inner disk boundary).}
  \label{o+cb.den}
\end{figure}

The density structure of the protoplanetary disk 
-- on which the subsequent radiative transfer (RT) simulations
and discussions are based on -- results from hydrodynamical
simulations in which the disk is assumed to be flat and non-self-gravitating.
The mutual gravitational interaction between the planet and the central star,
and the gravitational torques of the disk acting on planet and star
are included. 
The 3D density structure is Gaussian in the vertical
direction where for the scale height H(r) we assume a constant ratio
H/r=0.05. 
The radial density profile of the disk is shown in Fig.~\ref{o+cb.den}[A].

The mass of the star is assumed to be 1 ${\rm M}_{\sun}$, the mass of the planet is
1 Jupiter mass, and the mass of the disk is $0.05\,{\rm M}_{\sun}$. For the dust-to-gas
mass ratio we take the classical value of 1:100. The diameter of the disk is 104\,AU
which is comparable to the size of the disks seen in silhouette against the Orion Nebula
(McCaughrean \& O'Dell 1996).
The planet is located at a distance of 5.2\,AU from the star.
The structure of the spirals and the gap reaches an equilibrium after
about 150 orbits ($\approx 1800\,{\rm yrs}$). The specifics of the models and
the results, in particular the width and internal structure of the
gap, are described in detail in Kley (1999) and Kley et al.\ (2001).

\section{Radiative transfer simulations}\label{rttech}

We used the results from the hydrodynamical simulations
to calculate the dust continuum emission.
The RT has been performed with a 3D continuum code
which is based on the Monte Carlo method.
The RT is simulated self-consistently, taking into account both
the initial temperature of the dust due to viscous heating
and the additional energy input of the central star.
The dust density and temperature structure are defined on a grid
which is chosen to be identical with that of the hydrodynamical simulations.
Thus, any additional discretization error resulting from
the ``transfer'' of the density/temperature structure to the RT code is avoided.
For further information concerning the RT code, we refer to Wolf et al.\ (1999)
and Wolf \& Henning (2000).

In addition to the results from the hydrodynamical simulations,
we introduce the following RT parameters:
spherical dust grains consisting of ``astronomical'' silicates 
(optical data from Draine \& Lee 1984, radius 0.12\,${\rm \mu}$m); 
star: effective temperature $T_{\rm eff}=5500$\,K, luminosity $L = 1\,{\rm L}_{\sun}$; 
wavelength range for the simulation of the radiative transfer:
$0.03 \ldots 2000\,{\rm \mu}$m.
We should note that the  resulting temperature structure of the disk is in good agreement
with the temperature distribution resulting from 2D hydrodynamical disk models
with the same geometry.

\section{Images}\label{images}

\begin{figure*} 
  \epsscale{2.0}
  \plotone{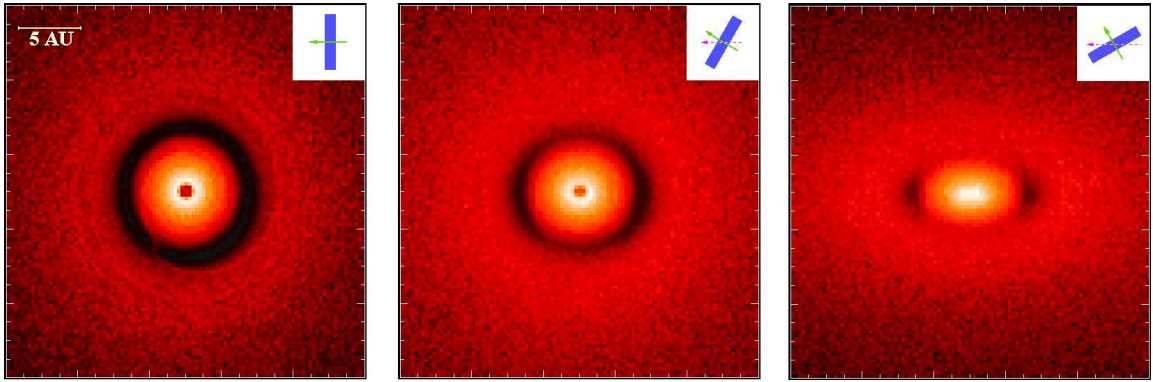}
  \caption{
  Images of the inner region (diameter 28\,AU; see Fig.~\ref{o+cb.den} 1[A]
  for the corresponding radial density profile) 
  of the circumstellar disk with an embedded planet of 1 Jupiter mass
  at a wavelength of $\lambda$ = 700\,$\mu$m
  and inclinations $i = 0^{\rm o}$ (face-on), $30^{\rm o}$, and $60^{\rm o}$ 
  (see symbols in the bottom row). 
  Each frame consists of $101 \times 101$ pixels. 
  The gap is clearly visible since the density drops by two orders of magnitudes
  in this region (optical depth at a wavelength of 700\,$\mu$m: $2\times 10^{-3}$).
  To increase the color dynamic range, the quantity $I_{\lambda}^\frac{1}{4}$ is shown. 
  The spiral density waves in the disks are weakly visible 
  at $i=0^{\rm o}$.
  The median specific intensity $I_{\lambda}$
  $\left[ {\rm W} \mu\rm m^{-1} {\rm sr}^{-1} {\rm m}^{-2} \right]$ per pixel amounts to
  1.74$ \cdot 10^{-23}$ ($i$=$0$\degr),
  1.68$ \cdot 10^{-23}$ ($i$=$30$\degr),
  9.93$ \cdot 10^{-24}$ ($i$=$60$\degr).
  See electronic edition for colored images.
  }
  \label{images12}
\end{figure*}

In Figure\,\ref{images12}, the images of the inner region (diameter 28\,AU) 
of the disk, seen under different inclination angles 
$i$ ($i= 0^{\rm o}$ - face-on, $30^{\rm o}$, and $60^{\rm o}$),
at $\lambda$=700\,$\mu$m are shown. Assuming a distance of 140\,pc of the object
(which is the distance to the nearby star-forming region in Taurus),
the pixel scale is 2\,mas/pixel.
The gap can be clearly seen -- even in the case of large inclinations
of the disk. The spiral density waves -- which are also caused by the motion 
of the planet -- are difficult to trace.

Due to the huge intensity peak caused by warm/hot dust
located very close to the star (1$\ldots$2\,AU), the outer regions of the disk
remain dark at near/mid-infrared wavelengths.  The gap located at 5\,AU can be hardly seen.
In contrast to this, in the sub-millimetre/millimetre wavelength 
range the brightness distribution is much smoother. This can be explained by the behaviour
of the Planck function $B_{\rm \lambda}(T)$. The ratio 
$B_{\rm \lambda}(T_1) / B_{\rm \lambda}(T_2)$ ($T_1 > T_2$) strongly decreases with
increasing wavelength. Thus, the flux ratio and the brightness contrast
between regions characterized by different dust temperatures ($T_1$ and $T_2$)
decreases with increasing wavelength $\lambda$.

\section{Simulations of observations with MIDI (VLTI) and ALMA}\label{simmial}

\begin{figure}[h]
  \epsscale{1.0}
  \plotone{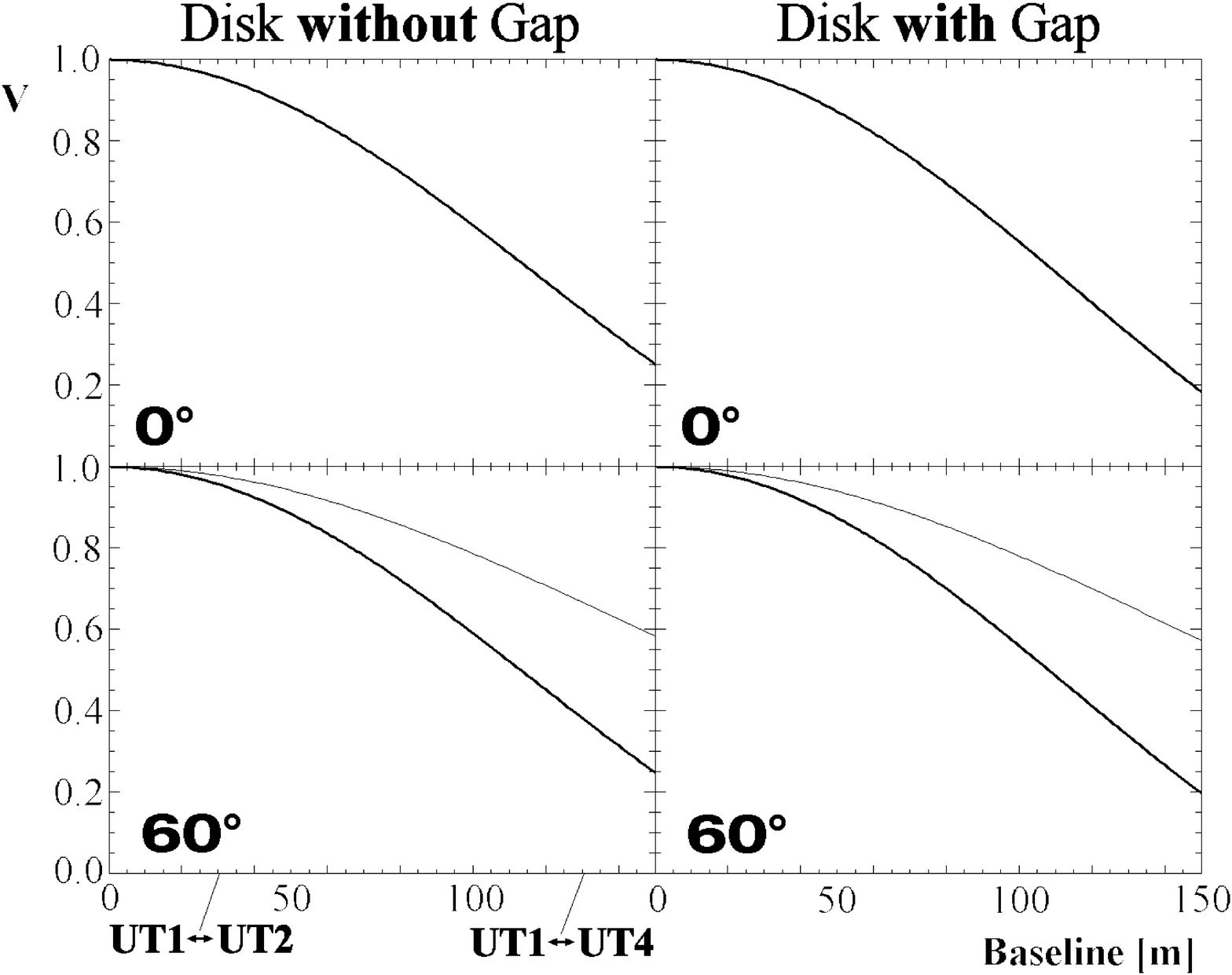}
  \caption{Normalized visibilities at $\lambda$=10\,$\mu$m for a disk with an inclination
    of $i=0^{\rm o}$ and $60^{\rm o}$ (right column; assumed distance: 140\,pc). 
    For comparison the same profiles in the u-v plane but for a disk without
    a gap (but with the same mass and inclinations) are also given (left column).
    The visibilities marked by the thick and the thin line
    are oriented perpendicular to each other in the u-v plane 
    (parallel to the major/minor axis of the ellipse resulting from 
    the projection of the disk onto the plane of the sky).}
  \label{simmidi}
\end{figure}

\begin{figure}[h]
  \epsscale{1.0}
  \plotone{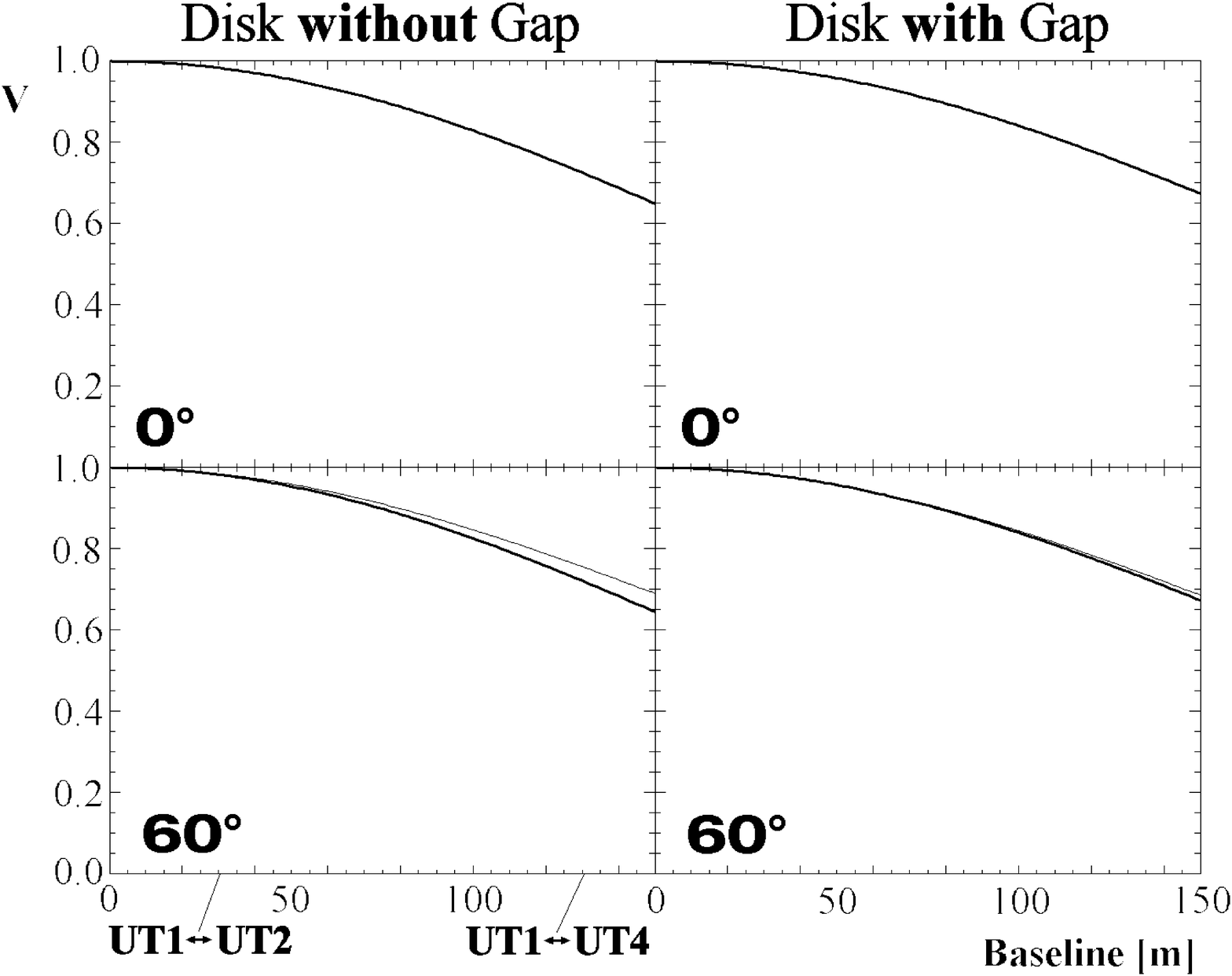}
  \caption{Normalized visibilities at $\lambda$=10\,$\mu$m for a disk with an inclination
    of $i=0^{\rm o}$ and $60^{\rm o}$ (right column; assumed distance: 140\,pc).
    In contrast to the visibilities shown in Fig.~\ref{simmidi}, a disk without accretion
    onto the central star is assumed here.}
  \label{simmidi2}
\end{figure}

In Figure\,\ref{simmidi} visibility curves at $\lambda$=10\,$\mu$m for two different
inclinations of the disk are shown (with/without gap). The goal of these simulations
is to check whether MIDI at the VLTI can be used to detect gaps.
In order to avoid sampling problems in the image plane which could affect
the calculation of the visibilities, the resolution of the images 
had to be increased by a factor of 4 along each coordinate axis 
(compared to Figure\,\ref{images12}). 
It was found that the visibilities at a given baseline differ by less than 5\%.
This is due to the fact that the density distribution in the innermost region
-- which dominates the 10\,$\mu$m flux -- 
is only negligibly affected by the presence of the planet at a distance of 5.2\,AU.
Taking into account uncertainties of ``real'' measurements
(e.g.\ with MIDI - the mid-infrared interferometric instrument for the VLTI),
a distinction between different disk models (with/without gap)
based on significant differences between the visibility profiles
is not possible even by a beam combination of the most distant telescopes of the VLTI.
To demonstrate the influence of the particular density profile of the disk,
in Fig.~\ref{simmidi2} the visibilities of a disk, with the same parameters 
as described in \S~\ref{model} but without accretion onto the central star,
are shown (see Fig.~\ref{o+cb.den}[B] for the radial density profile).
While again the presence of a gap has almost
no influence on the visibility, the visibility profile is flatter
compared to the visibilities shown in Fig.~\ref{simmidi}.
Thus, one could clearly distinguish a disk with accretion from a disk without.

\begin{figure}
  \begin{turn}{-90}
    \epsscale{1.0}
    \epsfig{file=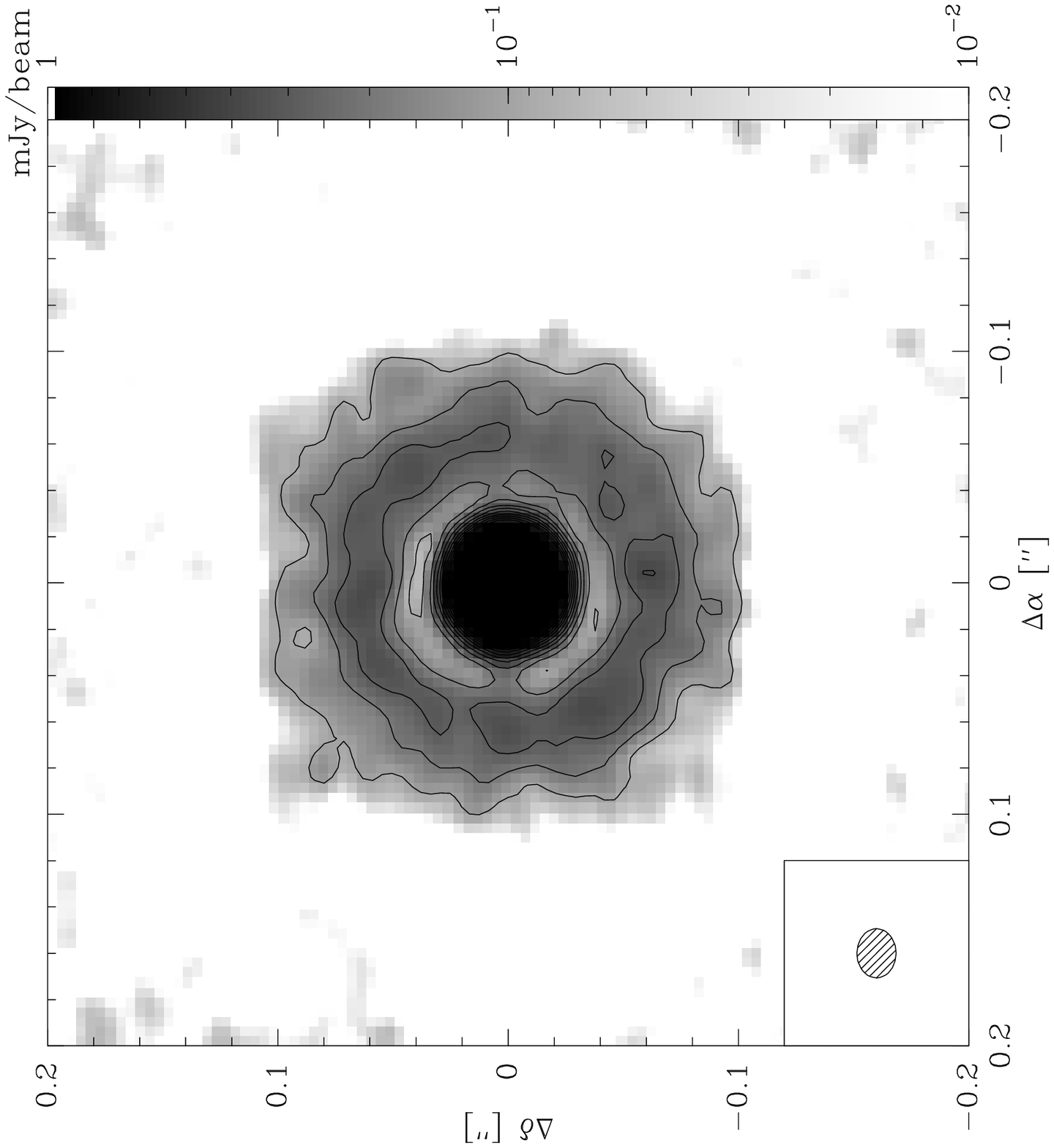,width=6cm}
  \end{turn}
  \caption{Reconstructed image of the disk seen face-on resulting 
    from a simulation of ALMA. The gap at an angular distance of
    37\,mas (5.2\,AU) from the star is clearly visible.  
    Wavelength: 700\,$\mu$m (428\,GHz);
    bandwidth: 8\,GHz;    total integration time: 4\,h;
    system temperature: 500\,K;
    phase noise: 30\degr;
    max.\ baseline: 10\,km.}
  \label{simalma}
\end{figure}

In contrast to a two-beam interferometer such as MIDI at the VLTI, 
ALMA (Atacama Large Millimeter Array) will combine signals of 64 antennas
with an aspired maximum baseline of 12 to 14\,km.
It will cover the submm/mm wavelength range ($\nu$=30$\ldots$900\,GHz).
Due to the large number of receivers and the resulting broad distribution
of baselines, a sufficient u-v plane coverage can be achieved after
a few hours of observation and image reconstruction will be possible.
In addition, and in contrast to the first instruments at the VLTI, the visibility
phase can be easily measured in the millimetre wavelength range.
In Figure\,\ref{simalma} a reconstructed image of the disk based 
on a simulation of ALMA observations is shown
(disk seen face-on, $\lambda$=700\,$\mu$m).
Even under consideration of the thermal noise caused by a system temperature
of $T_{\rm sys}$=500\,K, the gap is clearly visible.

\section{Conclusions} \label{concl}

Based on hydrodynamical simulations and subsequent 3D
continuum RT calculations we generated millimetre images 
of a circumstellar disk with an embedded planet surrounding a solar-type star.
The gap can be seen very clearly in the simulated
images but an extremely high angular resolution is required ($\approx$\,10\,mas).
Because of the extreme brightness contrast in the innermost region of the disk
in the near to mid-infrared, the gap can hardly be detected in this wavelength 
range with imaging/interferometric observations.

We found that it will be possible to distinguish between a disk with or 
without accretion onto the central star with MIDI, but the visibility
at 10\,$\mu$m is almost insensitive to the presence of a gap.
In contrast to this, the millimetre interferometer ALMA
which will become available in the near future 
will achieve this goal: it will provide the basis for the reconstruction of an image
of a gap. Thus, the search for massive protoplanets in circumstellar disks
can be based on the indication of a gap.
For this purpose, the longest possible baselines are required.

\acknowledgments

This research was supported by the DFG grant Kl\,650/1-1.
We appreciate the suggestions given by G.\ Bryden who was the referee
of this article.


\end{document}